# The clustering of warm and cool IRAS galaxies

R.G. Mann,[1,2] W. Saunders[3] and A.N. Taylor[1]
[1] *Institute for Astronomy, Department of Physics and Astronomy, University of Edinburgh, Blackford Hill, Edinburgh EH9 3HJ*
[2] *Present Address: Astronomy Unit, Queen Mary and Westfield College, Mile End Road, London E1 4NS*
[3] *Department of Physics, University of Oxford, Keble Road, Oxford OX1 3RH*

24 October 1995

**ABSTRACT**
We use a series of statistical techniques to compare the clustering of samples of IRAS galaxies selected on the basis of their far-infrared emission temperature, to see whether a temperature-dependent effect, such as might be produced by interaction-induced star formation, could be responsible for the increase in clustering strength with redshift in the QDOT redshift survey that has been reported by several authors.

The temperature-luminosity relation for IRAS galaxies means that warm and cool samples drawn from a flux-limited sample like QDOT will sample quite different volumes of space. To overcome this problem, and to distinguish truly temperature-dependent results from those depending directly on the volume of space sampled, we consider a pair of samples of warmer and cooler galaxies with matched redshift distributions, as well as pairs of samples selected using a simple temperature cut.

We find that the redshift space autocorrelation function of warm QDOT galaxies is significantly stronger than that of cool galaxies on large scales, but that this difference disappears when we come to consider the warmer and cooler samples with matched redshift distributions. A counts-in-cells analysis reveals no significant difference between the clustering of the warm and cool QDOT samples, while the use of a new, symmetric estimator reveals that the cross-correlations of warm and cool IRAS galaxies with Abell clusters do not differ significantly. A higher signal-to-noise test is provided by computing the projected cross-correlations of the matched samples with the parent two-dimensional catalogue from which QDOT is drawn and this does yield a marginal detection of greater large scale power for warmer galaxies. A direct comparison of the distributions of the warmer and cooler samples, using a new technique which tests the null hypothesis that they are drawn from the same population, reveals that the two classes of galaxy do cluster differently on small scales in redshift-space, while their $\xi(\sigma,\pi)$ plots suggest that the apparent concentration of more warm IRAS galaxies into richer environments reflects the fact they sample richer volumes of space within the QDOT survey, rather than illustrating a correlation between temperature and richness.

We conclude that there may be a temperature-dependent component to the observed increase in the clustering strength of QDOT galaxies with redshift, but that it is less important than a sampling effect, which reflects the local cosmography, rather than the physical properties of the galaxies and their environment. We discuss the implications of this work for the use of IRAS galaxies as probes of large-scale structure and for models accounting for their far-infrared emission by interaction-induced star formation.

**Key words:** Galaxies: clustering; cosmology: large-scale structure of Universe.

## 1 INTRODUCTION

IRAS offers the cosmologist excellent sky coverage, together with uniform flux calibration, good positional accuracy and insignificant Galactic absorption. These factors combine to make the distribution of IRAS galaxies a potentially important cosmological probe and one that has been employed by many authors in recent years.

Implicit in the use of IRAS galaxies as large-scale structure probes is an assumption about the homogeneity of their properties, at least in so far as these properties may be correlated with environment and, hence, with clustering strength. The overwhelming majority of galaxies observed by IRAS are spirals (e.g. de Jong et al. 1984), which are seen in preference to early-type galaxies because they contain more dust to re-radiate energy from galactic sources into the far-infrared. There are several mechanisms through which this can take place. Rowan-Robinson & Crawford (1989, hereafter RRC89) have modelled the IRAS spectra of galaxies as being composed of three components: a cool 'disc' component, peaking in the 100$\mu$m IRAS passband, due to the re-emission by interstellar dust grains of energy absorbed from the ambient galactic radiation field; a warm 'starburst' component which is produced by optically-thick dust clouds surrounding regions where massive stars are currently being formed and which peaks in the 60$\mu$m band; and a hot 'Seyfert' component, peaking in the 12 and 25$\mu$m bands and identified with emission from dust clouds surrounding compact power-law continuum sources. They argue that the IRAS spectra of their sample of 227 galaxies are well explained in terms of varying fractions of these three components, with the great majority of IRAS galaxies being dominated by the 'disc' and 'starburst' components. Similar models of IRAS emission have been considered by de Jong et al. (1984), Helou (1986), Telesco, Wolstencroft & Done (1988) and Bothun, Lonsdale & Rice (1989) amongst others.

The model presented by RRC89 is a simple one, but its basic notion of the far-infrared emission from IRAS galaxies being a combination of ambient cirrus emission and localised emission from active star-forming regions is probably correct. The relative contributions of these two components to the emission from a particular galaxy may be quantified by a dust emission temperature deduced from its 60 and 100$\mu$m fluxes. Saunders et al. (1990, hereafter S90) have determined the 60$\mu$m luminosity functions for warm and cool subsamples of the QDOT redshift survey of IRAS galaxies (Lawrence et al., in preparation): they define their subsamples on the basis of the RRC89 models, which predict that the emission from galaxies with dust temperatures above 36 K (classified as 'warm') is dominated by the 'starburst' component, while the 'disc' component dominates in ('cool') galaxies with emission temperatures below 36 K. They find that the overall luminosity function of IRAS galaxies is dominated by cool galaxies for luminosities below $\sim 5 \times 10^6\,h^{-2}\,L_\odot$ (the solar luminosity, $L_\odot$, is $L_\odot = 3.826 \times 10^{33}$ erg s$^{-1}$ and the Hubble constant, $H_0$, is written as $H_0 = 100\,h$ km s$^{-1}$ Mpc$^{-1}$) and by warm galaxies above that level.

An obvious consequence of this observed temperature-luminosity relation for IRAS galaxies is that there will be a radial gradient in the local mean galactic dust emission temperature in a sample of IRAS galaxies selected above a 60$\mu$m flux limit: as we show in Fig.1, the galaxies selected above the QDOT flux limit of 0.6 Jy become, on average, warmer with increasing redshift.

This radial temperature gradient in QDOT appears to be mirrored by a similar gradient in clustering strength. Feldman, Kaiser & Peacock (1994) find that the amplitude of the redshift-space power spectrum of the QDOT survey increases by up to a factor of 1.5 if they give more weight to more distant galaxies, and show that this is not due to a luminosity-dependence to the clustering strength: this is confirmed by the redshift-space correlation function analysis of Moore et al. (1994). Could this clustering strength gradient be connected with the temperature gradient, with the population of galaxies in QDOT changing with increasing redshift, becoming dominated by warmer, more strongly clustered galaxies? In particular, could this changing population become more concentrated in richer environments, as Mo, Peacock & Xia (1993) have shown that the cross-correlations between QDOT galaxies and Abell clusters strengthen with redshift?

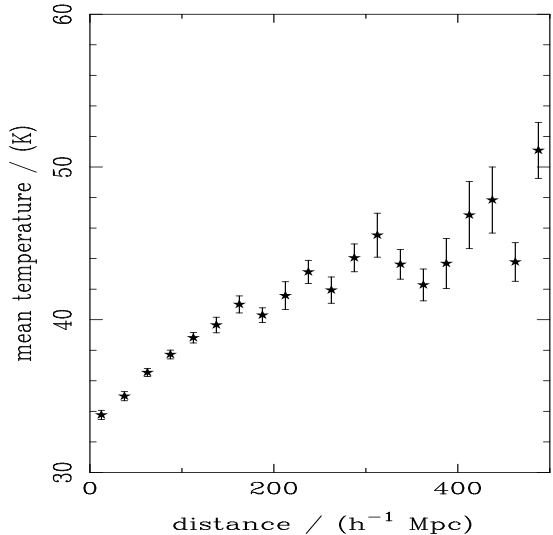

**Figure 1.** The radial temperature gradient in QDOT.

Simple notions of interaction-induced star formation may provide a link between the temperature of an IRAS galaxy and the richness of its environment and, hence, between the temperature and clustering strength gradients. There is a growing body of evidence linking the dynamical interactions between galaxies with the starburst phenomenon of enhanced star formation. Theoretical simulations of such events (e.g. Barnes & Hernquist 1991 and references therein) have shown how interactions can result in the gas in the interstellar media of the interacting galaxies losing angular momentum and the resulting infall of large quantities of gas could provide the conditions required to produce a burst of star formation. The relationship between interactions and mass infall is, however, a complicated one, depending on such factors as the closeness of approach and the orientation of the galaxies' axes of rotation relative to the direction of their orbital motion (e.g. Mihos et al. 1992). Observational evidence is equally forthcoming (see, for example, Sulentic, Keel & Telesco 1990 and references therein). Bushouse, Lamb & Werner (1988) and Xu & Sulentic (1991) have shown that samples of IRAS galaxies selected on the basis of morphological indications of interactions in the optical have higher far-infrared luminosities than samples of isolated galaxies showing no such features. Surace et al. (1993) find that while infrared properties alone are insufficient to distinguish clearly an *individual* interacting galaxy from an *individual* isolated galaxy with the same blue luminosity, due to the intrinsic dispersion in galactic properties, it is seen that *samples* of interacting galaxies have substantially warmer far-infrared colours than *samples* of isolated galaxies with the same distribution of blue luminosities.

One might naively expect that if it is the two-body process of galaxy interaction that is responsible for the starburst phenomenon of enhanced star formation, then starburst galaxies should be found preferentially in regions of high galactic density and, thus, one might expect there to be a relationship between the dust emission temperature of an IRAS galaxy and the richness of its environment, producing, in turn, a trend of increasing clustering strength with redshift, mirroring the temperature gradient. The high velocity dispersions in the cores of rich clusters will, however, prevent mergers taking place there and, so, limit the enhancement of clustering strength brought about by this effect, which we advance only as a plausible physical way of producing correlated temperature and clustering strength gradients. The widespread use of flux-limited samples of IRAS galaxies as cosmological probes does, however, strongly motivate a thorough investigation of the possible effect on large-scale structure studies of the differing $60\mu$m luminosity functions of samples of warm and cool IRAS galaxies, and that is our principal concern here.

We investigate the possible existence of such an effect by comparing the clustering strengths of subsamples of warm and cool IRAS galaxies in this paper: the dispersion in the temperature-$60\mu$m luminosity relation for IRAS galaxies means that this is not redundant in the light of previous searches for luminosity dependence in the clustering of QDOT galaxies.

The plan of the remainder of this paper is as follows. In Section 2 we describe the construction of warm and cool subsamples from QDOT and the determination of their redshift-space selection functions. The redshift-space autocorrelations of these subsamples are the subject of Section 3, while a complementary counts-in-cells analysis is given in Section 4, and, in Section 5, we probe the real-space clustering of the warm and cool samples through their projected cross-correlations with the galaxies of the QMW IRAS Galaxy Catalogue (QIQC: Rowan-Robinson et al. 1991). Section 6 makes a direct comparison of the redshift-space distributions of warm and cool IRAS galaxies, through the use of a new technique to test whether the warmer and cooler samples are drawn from the same population. We investigate the possibility that the warm galaxies are preferentially located in richer environments through studying the anisotropy in $\xi(\sigma,\pi)$ in Section 7, and the cross-correlations of warm and cool samples with Abell clusters in Section 8: the latter is conducted using a new cross-correlation function estimator, whose use we justify in the Appendix. Finally, in Section 9 we discuss the results of this project and present the conclusions we draw from them in Section 10.

## 2   DATA SAMPLES

### 2.1   The QDOT IRAS galaxy redshift survey

We consider warm and cool subsamples drawn from the QDOT IRAS galaxy redshift survey (Lawrence et al., in preparation). This comprises redshifts for a one-in-six sparse sample of galaxies drawn from the QIGC, limited to $|b| \geq 10°$ and above a $60\mu$m flux limit of 0.6 Jy, which marks the completeness limit of the IRAS Point Source Catalog (Chester, Beichman & Conrow 1987). We use the 1993 revision of the QDOT catalogue, in which the redshifts for ∼200 southern galaxies afflicted by a wavelength calibration error in earlier versions of the catalogue have been corrected.

### 2.2   The Mask

The standard QIGC/QDOT mask excludes from analysis that portion of the sky (about 4 per cent) missed by IRAS, plus the region $|b| \leq 10°$ and various other 'lune bins' ($1° \times 1°$ regions defined in ecliptic coordinates) where high source density confuses discrimination between Galactic and extragalactic sources. Our purposes, however, necessitate the exclusion of further lune bins, as we require accurate temperature determination on the basis of 60 and $100\mu$m fluxes.

Not all QDOT galaxies have confirmed detections at $100\mu$m and the variation of the $100\mu$m Galactic cirrus emission (e.g. Low et al. 1984) across the sky results in a corresponding variation in the upper limits assigned to sources in the absence of confirmed $100\mu$m detections. This situation may be greatly ameliorated by the exclusion of lune bins with low $100\mu$m upper limits and we implement that here, extending our masked region to exclude from our analysis those lune bins where the $100\mu$m background flux due to Galactic cirrus emission exceeds 15 MJy per steradian: with our mask thus extended, our QDOT sample covers 8.84 steradian (i.e. about 70 per cent of the sky) and comprises 2053 galaxies. A number of galaxies with $100\mu$m upper limits remain even after the mask is extended and we discuss what should be done with them in the next subsection.

### 2.3   Temperature assignment and subsample selection

Our assignment of dust emission temperatures to galaxies follows the method described by S90. We assume that the far-infrared spectrum of each galaxy is well described by a single-temperature Planck function multiplied by a $\lambda^{-1}$ emissivity, as often considered appropriate for dust: the modelling by RRC89 shows this to be a reasonable approximation for both 'disc'- and 'starburst'-dominated emission. For each source we fit a single-temperature $S_\nu \propto \nu B_\nu(T_{\rm obs})$ curve [where $S_\nu$ is the flux at frequency $\nu$ and $B_\nu(T)$ is the Planck function for temperature $T$] to the spectrum, such that its convolution with the response curves for the 60 and 100 $\mu$m IRAS detectors (Beichman et al. 1988) give the observed fluxes in those passbands. The emission temperature, $T_{\rm em}$, is then given by $T_{\rm em} = T_{\rm obs}(1+z)$ for a galaxy with redshift $z$.

We follow S90 in adopting a temperature of 36 K to mark the division between the warm and cool samples. This is motivated by the RRC89 models, but is also supported by the work of Bothun et al. (1989), who show that the mean temperatures of their 'normal' and 'active' ('disc' and 'starburst/Seyfert', respectively, in the parlance of RRC89) samples of UGC galaxies observed by IRAS are 35 and 38 K, respectively: it must be emphasised, however, that the intrinsic dispersion in galactic properties means that this temperature cannot be regarded as a sharp cut-off, nor should the warm and cool subsamples it defines be thought of as comprising totally distinct physical species of galaxy.

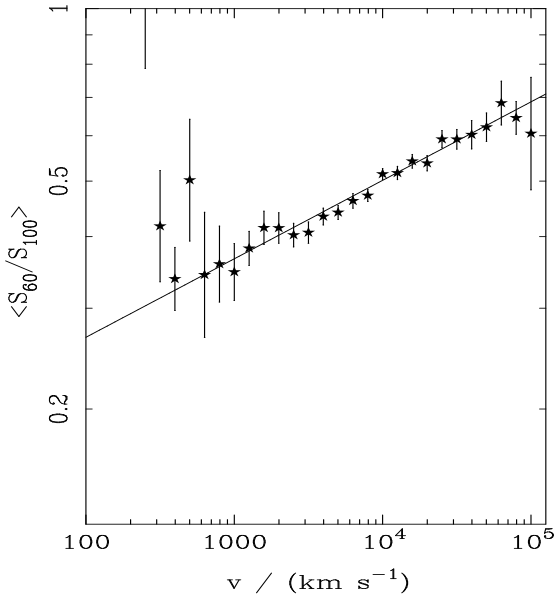

**Figure 2.** The variation in the median $S_{60}/S_{100}$ ratios in radial velocity shells. The solid line shows the power law fit used in the construction of the warmer and cooler matched $n(z)$ samples.

There remain some 275 galaxies within the unmasked region of the sky with $100\mu$m upper limits. If the $100\mu$m upper limit for a particular galaxy is sufficiently low compared to its observed $60\mu$m flux, then we may safely classify it as warm, according to our temperature criterion, but 161 galaxies remain with uncertain temperature classifications in the unmasked region of our QDOT sample after accounting for this. For these galaxies we take the $100\mu$m upper limit as a detection. This is not so questionable an action as at first it might seem, as the great majority (141 out of 161) of these sources were detected at $100\mu$m in at least one IRAS scan and the upper limit quoted is the flux measured in this unconfirmed detection. We may be sure that the 94 galaxies in our QDOT sample classified as warm by this procedure must truly belong to that category, as their $100\mu$m fluxes can only be lower and, hence, their temperatures can only be higher than we have imagined, leaving us with a total of 57 galaxies with doubtful cool classifications (since 10 galaxies not detected at $100\mu$m were classified as cool on the basis of their upper limits in that passband). Some of these, no doubt, belong in the warm sample, but we shall assume that this has no bearing on our clustering analysis. We may test the validity of this assumption by analysing samples selected above a higher $60\mu$m threshold, which will have fewer uncertain cool classifications: only 13 problem cool sources remain in our QDOT sample when the $60\mu$m flux threshold is increased to 0.7 Jy, for example.

The radial temperature gradient in QDOT means that the warm and cool subsamples created this way have very different redshift distributions, making it difficult to distinguish temperature-dependent effects from those which depend directly on redshift. To circumvent this problem, we have constructed a further pair of subsamples in such a way as to survey the same volume with each. We have done this by splitting QDOT into equal sized warmer and cooler subsamples according to the median temperature as a function of distance:

(i) We split the sample into recessional velocity shells of width 0.1 dex.

(ii) We find the median $S_{60}/S_{100}$ ratio for galaxies within each shell, where $S_{60}$ and $S_{100}$ are, respectively, the IRAS fluxes in the $60\mu$m and $100\mu$m passbands.

(iii) We fit this median ratio as a power law with distance, with the result $\langle S_{60}/S_{100} \rangle = 0.142 v^{0.137}$, where $v$ is the recessional velocity in km s$^{-1}$. The power law fit for temperature versus distance is shown in Fig. 2 and appears to be an adequate model, in that the $\chi^2$ between numbers of warmer and cooler galaxies is 21 for 25 degrees of freedom, assuming Poisson statistics.

(iv) We find whether each galaxy is warmer or cooler than the median ratio for its distance and put it into the warmer or cooler subsample accordingly: these samples have, by construction, almost identical redshift distributions.

**Table 1.** Properties of principal galaxy samples

| sample | no. of galaxies | mean temp. (K) |
|---|---|---|
| warm, temp-selected | 1233 | 41.84 |
| cool, temp-selected | 820 | 32.04 |
| warmer, matched $n(z)$ | 1002 | 41.95 |
| cooler, matched $n(z)$ | 1010 | 34.47 |
| warm, temp, $S_{60} > 0.7$ Jy | 997 | 42.19 |
| cool, temp, $S_{60} > 0.7$ Jy | 635 | 32.56 |

In Table 1 we list the number of galaxies in each of our principal IRAS galaxy samples, together with their mean temperatures. In Fig. 3 we plot the temperature distributions of the two pairs of samples selected above a $60\mu$m flux limit of 0.6 Jy: note the different temperature scale in Fig. 3(c). This figure shows the large degree of overlap between the warm (cool) temperature-selected sample and the warmer (cooler) matched $n(z)$ sample: the fact that the temperature distributions of the corresponding members of these pairs of samples are so similar, coupled with the negligible difference between the redshift distributions of the two matched $n(z)$ samples, means that we can combine results from these two pairs of samples to distinguish between temperature-dependent effects and sampling or luminosity-dependent effects that depend more directly on redshift.

### 2.4 Selection functions

The radial decline in the number density of galaxies in a flux-limited sample is quantified by its selection function, $\phi(r)$, which we define to be the expected number density of galaxies at distance $r$ lying above the sample's flux limit in the absence of clustering.

Throughout this paper we correct radial velocities to the centroid of the Local Group by adding to the observed heliocentric velocity of a galaxy with Galactic longitude $l$ and latitude $b$ the correction term 300 $\sin(l)\cos(b)$ km s$^{-1}$ and assume an Einstein - de Sitter universe ($\Omega_0 = 1, \Lambda = 0$).

Details of the methods used in the computation of the selection functions for our sample will be presented elsewhere (Saunders, in preparation) and we provide only a summary here. We use both parametric and non-parametric forms for the selection function, which we determine by the following procedure:

(i) The galaxies in a given sample are binned, to generate a redshift distribution, $n(r_j)$, which gives the number of galaxies in the bin centred on distance $r_j$, measured in $h^{-1}$ Mpc.

(ii) A naive estimate of the selection function, $\phi_1(r_j)$, is then obtained by dividing $n(r_j)$ by the volume of the bin, $V(r_j)$.

(iii) To account for the clustering in the galaxy distribution, we divide $\phi_1(r_j)$ by $\rho(r_j)$, which is the maximum-likelihood radial density estimator described by S90.

(iv) The quantity so obtained [call it $\phi_2(r_j)$] is, however, still not an adequate estimate of the selection function, since dividing through by $\rho(r_j)$ will remove the effect of number density evolution, which S90 showed to be important for QDOT. The final estimator for the selection function is obtained, therefore, by multiplying $\phi_2(r_j)$ by the evolution factor, $\bar{\rho}(r_j)$: note that this correction divides out any dependence of the selection function on assumed K-corrections or cosmological model. Combining these factors gives our final non-parametric estimate of the selection function as

$$\phi(r_j) = \frac{n(r_j)\bar{\rho}(r_j)}{V(r_j)\rho(r_j)}, \qquad (1)$$

and this is then normalised as discussed below.

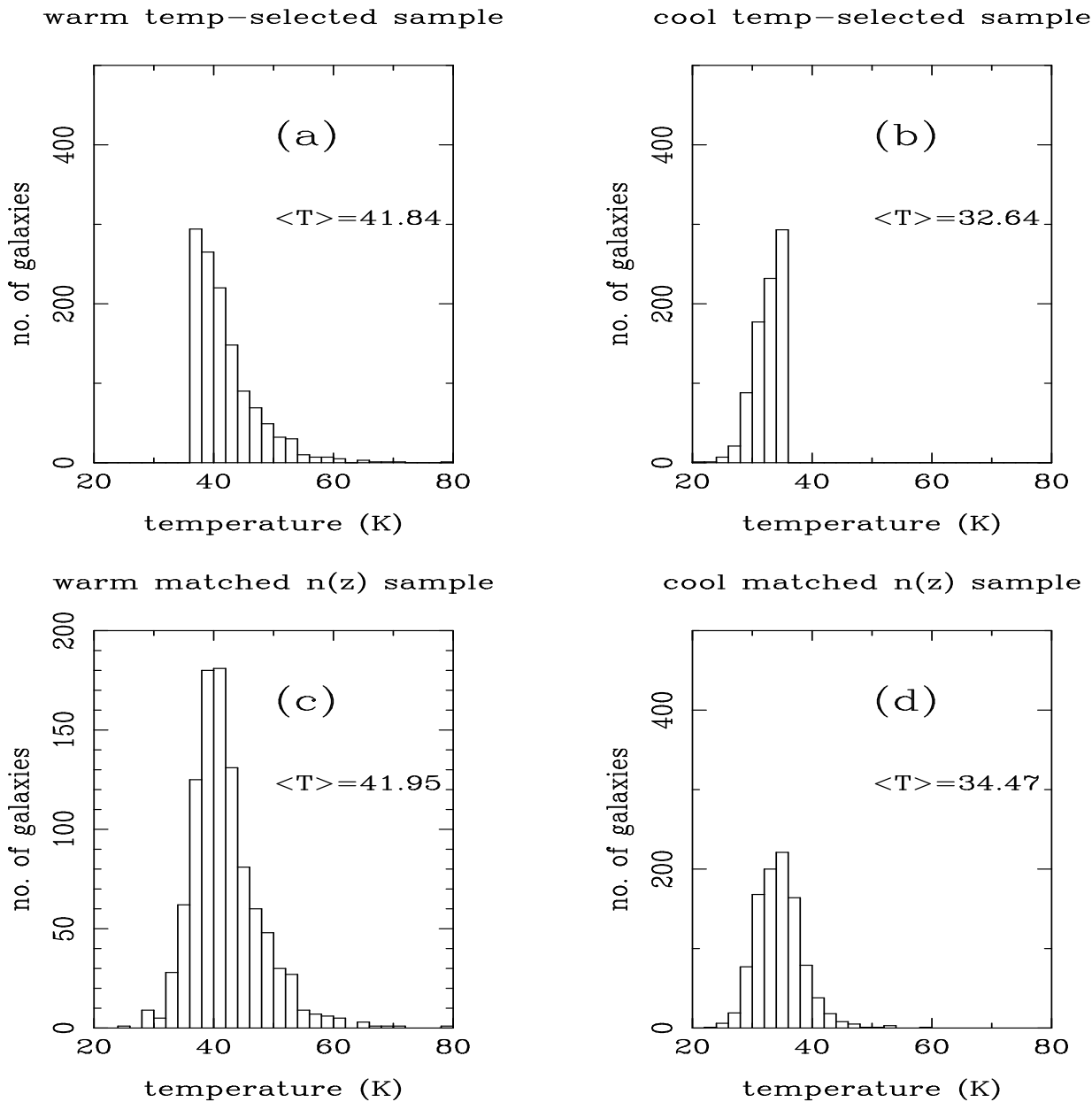

**Figure 3.** The temperature distributions and mean temperatures for the four QDOT subsamples selected above a $60\mu$m flux limit of 0.6 Jy: (a) warm temperature-selected sample; (b) cool temperature-selected sample; (c) warmer matched $n(z)$ sample; and (d) cooler matched $n(z)$ sample. Note the different scale in (c).

**Table 2.** Selection function parameters for the samples listed in Table 1

| sample | $10^4 \phi_*$ | $\alpha$ | $\beta$ | $\gamma$ | $\log_{10} r_*$ |
|---|---|---|---|---|---|
| QDOT | 22.0 | 1.61 | 3.90 | 1.64 | 1.81 |
| warm, temp-selected | 1.80 | 1.91 | 4.11 | 1.99 | 2.09 |
| cool, temp-selected | 10.4 | 1.87 | 4.61 | 2.42 | 1.77 |
| matched $n(z)$ samples | 11.0 | 1.61 | 3.90 | 1.64 | 1.81 |
| warm, $S_{60} > 0.7$ Jy | 1.88 | 1.91 | 4.07 | 1.94 | 2.06 |
| cool, $S_{60} > 0.7$ Jy | 9.37 | 2.05 | 4.63 | 2.52 | 1.74 |

It is found that the selection function so computed closely approximates a double power law (Fig. 4), so a convenient parametric form for the selection function is

$$\phi(\Delta) = \phi_* \frac{10^{(1-\alpha)\Delta}}{(1 + 10^{\gamma\Delta})^{\beta/\gamma}}, \qquad (2)$$

where $\Delta = \log_{10}(r/r_*)$, for a galaxy at a distance, $r$, measured in units of $h^{-1}$ Mpc, and $\phi_*$, $\alpha$, $r_*$, $\gamma$ and $\beta$ are parameters describing, respectively, the amplitude, nearby slope, break position and breadth, and distant slope of the selection function.

The parameters $\alpha$, $r_*$, $\gamma$ and $\beta$ are determined by maximising the likelihood quantity

$$\mathcal{L}_\phi = \text{const} \times \prod_{\text{sources}} \frac{\bar{\rho}(r_i)/\phi(r_i)}{\sum_{\text{gals with } r_k < r_{max,i}} \bar{\rho}(r_k)/\phi(r_k)} \qquad (3)$$

where $r_{max,i}$ is distance at which each galaxy, of given observed luminosity, would drop below the flux limit, and $\bar{\rho}(r_k)$ reflects the level of evolution found as above.

The amplitude, $\phi_*$, is determined by demanding that the $J_3$-weighted integral over the selection function equal the $J_3$-weighted sum over the sample:

$$\Omega \int \frac{dV}{dr} \frac{\phi(r)}{(1 + 4\pi J_3 \phi(r))} dr = \sum_{\text{sources}} \frac{1}{(1 + 4\pi J_3 \phi(r_j))} \qquad (4)$$

where $r_j$ is the distance to the $j$-th source; this sum is easily solved by iteration.

In Table 2 we give the values for these quantities appropriate to our principal samples and in Fig. 4 we show the selection functions derived by these methods for the warm and cool subsamples derived from QDOT with a 60 $\mu$m cut at 0.6 Jy.

## 3 REDSHIFT-SPACE AUTOCORRELATION FUNCTIONS

### 3.1 Method

We follow Hamilton (1993) in using the estimator

$$1 + \xi_{\text{AA}}(s) = \frac{D_\text{A} D_\text{A}(s) \cdot R_\text{A} R_\text{A}(s)}{[D_\text{A} R_\text{A}(s)]^2}, \qquad (5)$$

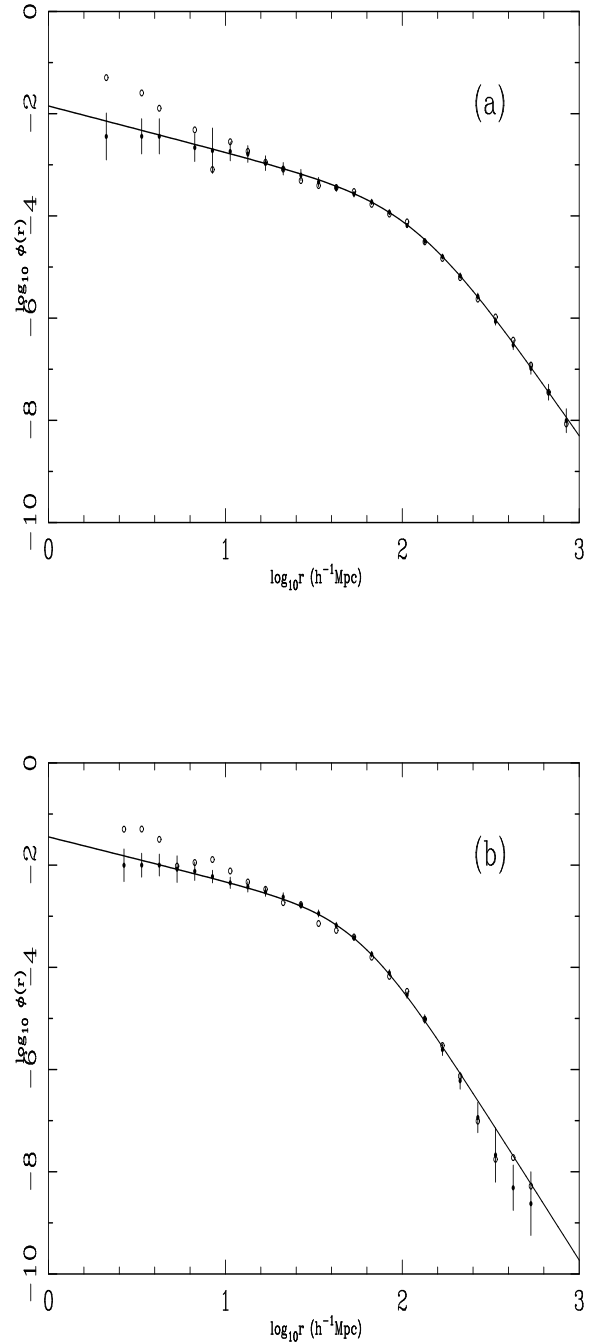

**Figure 4.** Selection functions for subsamples drawn from QDOT with a 0.6 Jy flux limit at 60 $\mu$m: (a) warm and (b) cool. The solid line shows the parametric fit, while the solid and empty circles show, respectively, the non-parametric selection function, $\phi(r)$, and the naive estimate, $\phi_1(r)$.

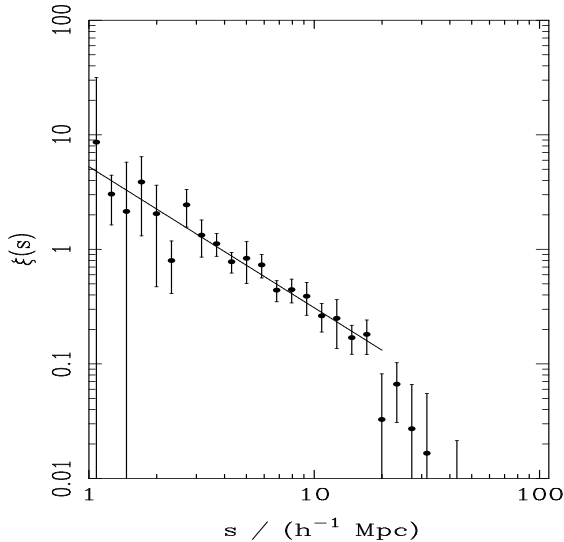

**Figure 5.** The redshift-space autocorrelation function for our QDOT sample. The solid line shows the best fit power law ($s_0 = 3.86 \pm 0.33\,h^{-1}$ Mpc, $\gamma = 1.23 \pm 0.08$) over the range from which it was determined

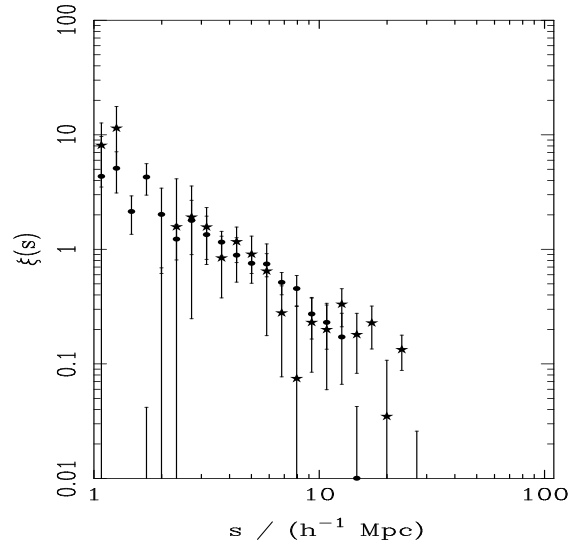

**Figure 6.** The redshift-space autocorrelation functions for the warm (stars) and cool (circles) temperature-selected samples.

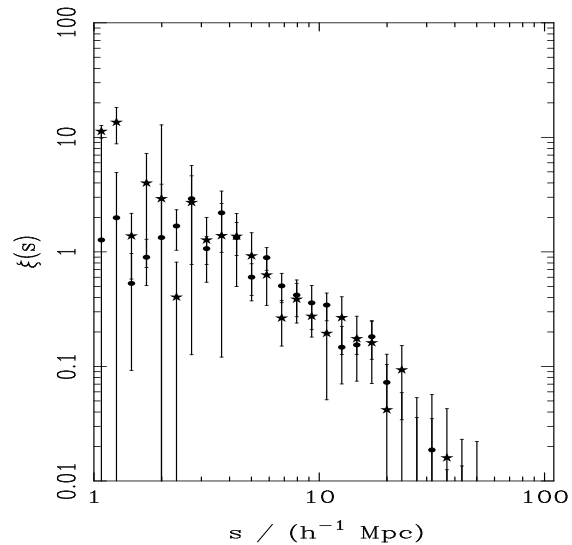

**Figure 7.** The redshift-space autocorrelation functions for the warmer (stars) and cooler (circles) matched $n(z)$ samples.

to calculate the redshift-space autocorrelation function, $\xi_{AA}(s)$, of galaxies of species A, where $D_A D_A(s)$ is the weighted pair count of $A$ galaxies whose separation places them in a bin centred on $s$, $R_A R_A(s)$ is the corresponding weighted pair count for a random catalogue with the same sky coverage and selection function as the galaxy sample and $D_A R_A(s)$ is the weighted count of cross-pairs between the galaxy and random catalogues. Hamilton (1993) has shown that this estimator is much less sensitive to uncertainty in the assumed mean number density of galaxies than the standard estimator

$$1 + \xi_{AA}(s) = \frac{D_A D_A(s)}{D_A R_A(s)} \cdot \frac{n_{R_A}}{n_A}, \qquad (6)$$

where $n_A, n_{R_A}$ are, respectively, the estimated mean number densities of the galaxy and random catalogues. Loveday et al. (1995) found that the correlation function of the Stromlo-APM survey computed using equation (5) is in good agreement with that for a volume-limited sample of the survey, while equation (6) yields spurious power on large scales, which they explain as resulting from slight differences between the radial density distributions of the galaxy and random catalogues. We also find that the standard estimator produces more power on large scales, and, sharing the interpretation of this advanced by Loveday et al. (1995), we favour the use of Hamilton's estimator.

To optimise signal-to-noise, it is necessary to weight the galaxies in some way that accounts for the variation in the mean galaxy number density with redshift. The simplest method (Davis & Peebles 1983) is to weight each galaxy by the reciprocal of the selection function at its redshift, so that a given volume of space is weighted by the number of galaxies expected to occupy it in a volume-limited sample. This prescription gives high weights to the most distant galaxies in a flux-limited sample, so a distance limit must be imposed if the resultant weighted data-data pair count is not to be blighted by shot noise. An alternative scheme (Efstathiou 1988, Loveday et al. 1992, Saunders, Rowan-Robinson & Lawrence 1992), and one which avoids the loss of information from galaxies beyond the distance limit, is to assign, when computing the pair count of galaxies with separation $s$, the weight

$$w(r) = 1/[1 + 4\pi f \phi(r) J_3(s)], \qquad (7)$$

to a galaxy at distance $r$, where $\phi(r)$ is the selection function at that distance, $f$ is the sparse-sampling factor (i.e. $f = 1$ for a fully-sampled survey and $f = 1/6$ for QDOT) and

$$J_3(s) = \int_0^s \xi(x) x^2 \mathrm{d}x. \tag{8}$$

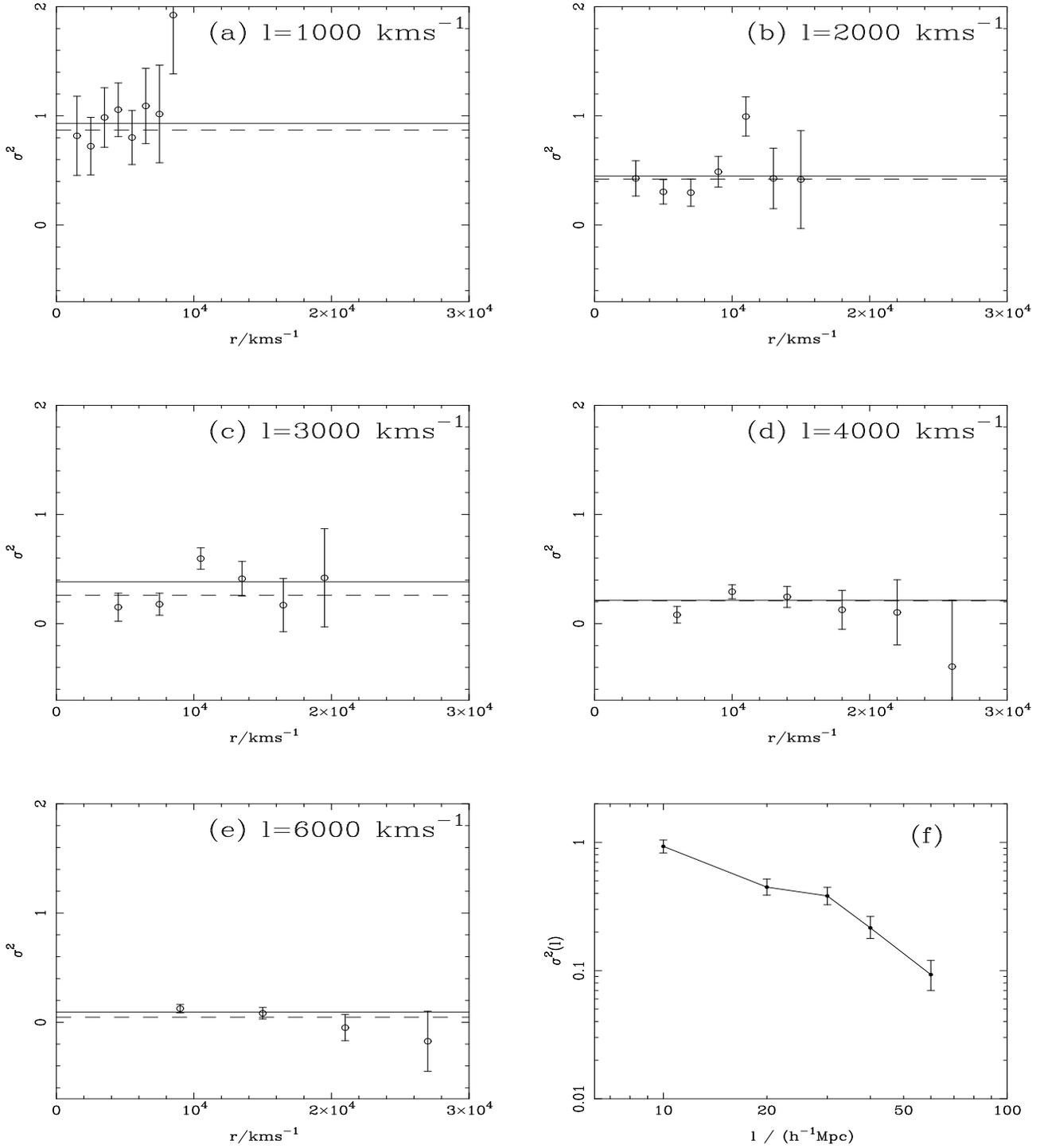

**Figure 8.** Counts-in-cells results for our full QDOT sample, to a depth of 300 $h^{-1}$ Mpc. In (a) – (e) we show results for $\sigma^2(l)$ in different radial shells for cells of size: (a) 1000 km s$^{-1}$; (b) 2000 km s$^{-1}$; (c) 3000 km s$^{-1}$; (d) 4000 km s$^{-1}$; and (e) 6000 km s$^{-1}$. Error bars are as estimated in the manner described by Efstathiou et al. (1990). In each plot the solid line denotes the maximum likelihood estimate of $\sigma^2(l)$ and the dashed line that from Efstathiou et al. (1990). In (f) we plot the maximum likelihood estimate of $\sigma^2(l)$ as a function of $l$.

It can be shown (e.g. Loveday et al. 1995) that this minimises the variance in the correlation function on large scales where the 'cluster model' of Peebles (1980) provides a good estimate of the uncertainty in the correlation function. The utility of this weighting scheme is clear: for small separations, where $4\pi f \phi J_3 \gg 1$ and where the contributions to the pair count variance are dominated by clustering rather than the discreteness of galaxies, the galaxies are weighted by the reciprocal of the selection function and, hence, equal volumes are weighted equally; while, on large scales, galaxies are given equal weight, thus reducing the problem of shot noise from a few, highly-weighted, distant galaxies.

For our model $J_3(s)$, we used the integral over the real space correlation function of QDOT galaxies determined by Saunders et al. (1992), multiplied by the linear theory redshift-space correction factor, $F = [1 + (2\Omega_0^{0.6}/3b) + (\Omega_0^{1.2}/5b^2)]$, derived by Kaiser (1987): Saunders et al. (1992) suggest that a value of $F = 1.57 \pm 0.32$ is appropriate for QDOT, although, in fact, the correlation function results are quite insensitive to the exact $J_3$ model used. We used 20,000 particle random catalogues throughout: tests showed that this was sufficient, as negligible difference was found when the number of random particles was increased to 100,000. We compute error bars from the scatter between correlation functions measured in octants.

### 3.2 Results

We first computed the autocorrelation function of our QDOT sample, which is shown in Fig. 5. The best fit power law of the form $\xi(s) = (s_0/s)^\gamma$ over the range $1 \le s \le 20\ h^{-1}$ Mpc was found to be $s_0 = 3.86 \pm 0.33\ h^{-1}$ Mpc, $\gamma = 1.23 \pm 0.08$, which is in good agreement with the results of Moore et al. (1994), who found $s_0 = 3.87 \pm 0.32\ h^{-1}$ Mpc, $\gamma = 1.11 \pm 0.09$ for their QDOT sample, which has a slightly different mask. In Fig. 6 we plot the autocorrelation functions of the warm and cool temperature-selected samples: very similar results were obtained from warm and cool samples selected above a $60\mu$m flux limit of 0.7 Jy and when, instead, we used the non-parametric selection function, justifying our procedure for treating the galaxies with $100\mu$m upper limits and our choice of parametric form for the selection function. From Fig. 6, we see that the warm sample is clearly more strongly clustered on large scales: there is a $4\sigma$ difference in the correlation functions over the range $17 \lesssim s \lesssim 50\ h^{-1}$ Mpc. This difference vanishes, however, when we come to the autocorrelation functions of the warmer and cooler samples with matched redshift distributions (Fig. 7), which do not differ by as much as $2\sigma$ in the full range $1 \lesssim s \lesssim 100\ h^{-1}$ Mpc: this strongly suggests that the principal cause of the difference in clustering strength seen in Fig. 6 is a sampling effect, resulting from the different redshift distributions of the warm and cool samples, rather than a temperature-dependent one.

### 4 CELL COUNT VARIANCES

An alternative statistic for quantifying the clustering of a sample of galaxies is the cell count variance (Peebles 1980;

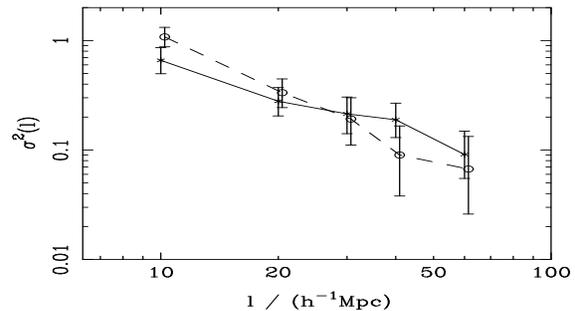

**Figure 9.** Counts-in-cells results for warm and cool QDOT samples, limited to a depth of 300 $h^{-1}$ Mpc: the dashed line denotes the variation of $\sigma^2(l)$ as a function of $l$ for the cool subsample and the solid line that for the warm subsample. The data points for the cool sample are slightly displaced horizontally for the sake of clarity and error bars are estimated in the manner described by Efstathiou et al. (1990).

Efstathiou et al. 1990; Loveday et al. 1992). The principal advantage that counts-in-cells analysis has over analyses using correlation functions is that, in principle, by considering the variance between cells at the same redshift, no account need be taken of the decline in the mean density with redshift in a flux-limited sample. The counts-in-cells analysis of this subsection is, therefore, intended to complement the correlation function analysis discussed above and, in particular, to ensure that the results obtained above reflect the real clustering of the IRAS galaxy samples and are not artefacts produced by their differing selection functions. The method of counts-in-cells analysis is described by Efstathiou et al. (1990) and we refer the reader to that paper for a full discussion of it. Oliver et al. (1995) have recently pointed out that the method of Efstathiou et al. (1990) produces biased estimates of cell count variances, as a result of its assumption that the galaxy selection function is constant throughout a particular cell, whereas, in fact, the decline in the selection function means that galaxies are more likely to be located towards the inner edge of any cell. This bias can be problematic when observational cell count variances are being compared with theoretical predictions, but it is of little concern to us here, as the differential bias caused by the difference in the selection functions of our warm and cool samples is likely to be much smaller than the uncertainties in our estimations of the cell count variances, given the relatively small size of our galaxy samples: we have not, therefore, included the volume-limiting correction of Oliver et al. (1995) in the counts-in-cells analysis whose results are presented here.

The first results of our counts-in-cells analysis are shown in Fig. 8, where we present results for our full QDOT sam-ple. This figure compares our results with those obtained by Efstathiou et al. (1990) and shows that the presence of erroneous redshifts in the earlier version of the QDOT catalogue used in that paper had no bearing on the counts-in-cells analysis performed there. This figure also shows that the values of $\sigma^2(l)$ for a given $l$ vary greatly between radial shells and each has a large uncertainty associated with it. The situation is even worse once we come to split QDOT into warm and cool samples or reduce the galaxy numbers by considering higher $60\mu$m flux limits. This leaves us very reliant on the ability of the maximum likelihood mechanism to extract the correct solution for $\sigma^2(l)$ as a function of $l$. We may try to facilitate this by excluding those shells where the counts are low, but we must inevitably contend with large uncertainties resulting from our small sample sizes.

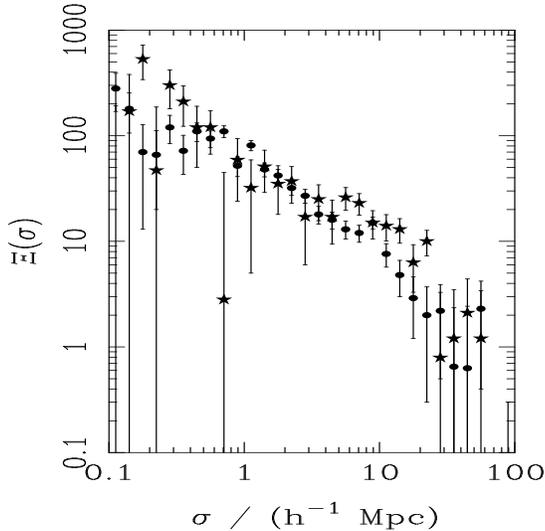

**Figure 10.** The projected cross-correlations of warm (stars) and cool (circles) temperature-selected samples of QDOT galaxies with the QIGC.

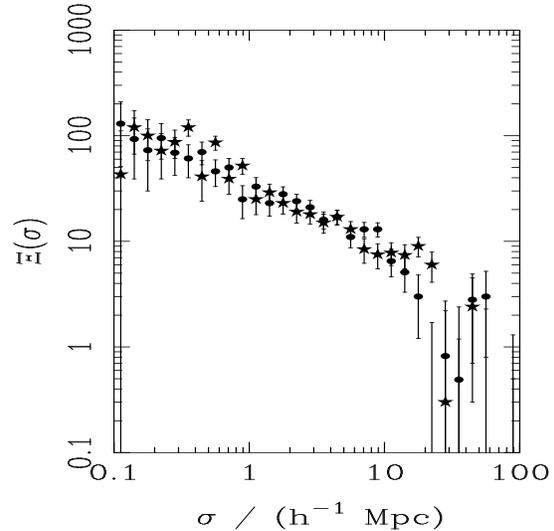

**Figure 11.** The projected cross-correlations of warmer (stars) and cooler (circles) matched samples of QDOT galaxies with the QIGC.

In Fig. 9 we present results for $\sigma^2(l)$ as a function of $l$ for the warm and cool temperature-selected subsamples limited to a depth of 300 $h^{-1}$ Mpc. This figure shows that the $\sigma^2(l) - l$ relations for warm and cool subsamples of QDOT galaxies are consistent within their errors. Very similar results are obtained using QDOT galaxies out to different depths, selected above different $60\mu$m flux limits and for the warmer and cooler samples with matched redshift distributions.

## 5 PROJECTED CROSS-CORRELATIONS WITH QIGC

The redshift-space tests carried out above suffer from poor signal-to-noise. In order to improve signal-to-noise, we have used the cross-correlation technique of Saunders et al. (1992) between warm and cool temperature-selected redshift subsamples and the whole 2-D QIGC parent catalogue from which they were drawn. This measures $\Xi(\sigma)$, the integral of $\xi(r)$ along a ray with projected separation $\sigma$: we refer the reader to Saunders et al. (1992) for a full discussion of the method employed here to estimate $\Xi(\sigma)$. The results shown in Fig. 10 show clear evidence for enhanced large scale clustering for the warm subsample as compared with the cool sample: as discussed by Saunders et al. (1992), the absolute level of $\Xi(\sigma)$ on large scales may not be accurately estimated by this method, due to the breakdown of several assumptions made in its derivation (see Saunders et al. 1992 for more details), but the difference between the values for the different subsamples should just result from shot noise if the subsamples were drawn from the same population.

To assess whether this difference might just result from the different volumes surveyed by the temperature-selected samples, we then repeat the cross-correlation analysis with each of the matched subsamples against QIGC. The result is Fig. 11 and, although much weaker, it appears that the evidence for stronger large-scale clustering of the warmer galaxies remains: since the $n(z)$ distributions are now essentially identical, the difference cannot be due to volume effects. We have, therefore, a marginal detection, at the few sigma level, of a real difference in the large-scale clustering of warm and cool IRAS galaxies.

## 6 DIFFERENTIAL DENSITY MAPS OF IRAS GALAXIES

In this Section we test the null hypothesis that the density pattern mapped by each of the warm and cool galaxy subsamples is the same: i.e. that they are random samples of the same underlying density field. Variations between the two fields should be accounted for by Poisson statistics.

To test this, we construct the reduced $\chi^2$ statistic for two populations of density $\rho$ and $\rho'$, with shot noise variances $\sigma_\rho^2 = \lambda$;

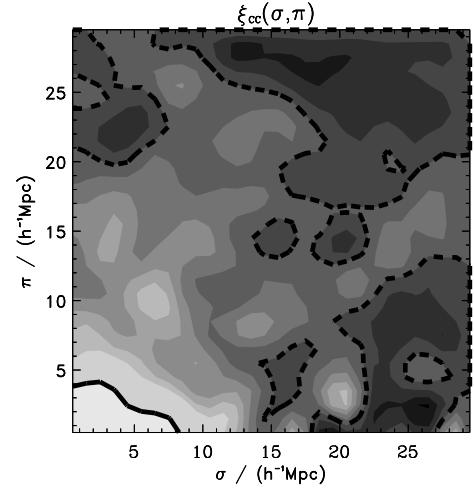

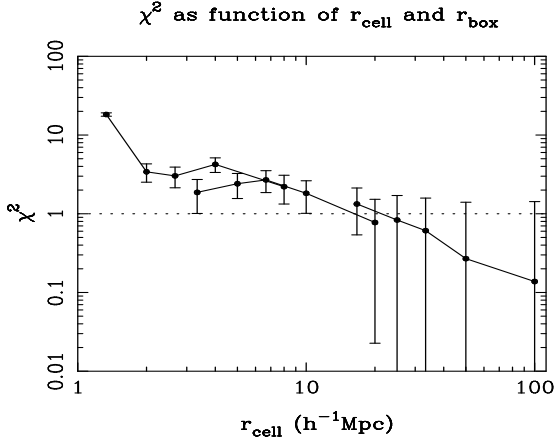

**Figure 12.** The $\chi^2$ density difference for the warm and cool subsamples of IRAS galaxies, as a function of cell size $r_{\rm cell}$.

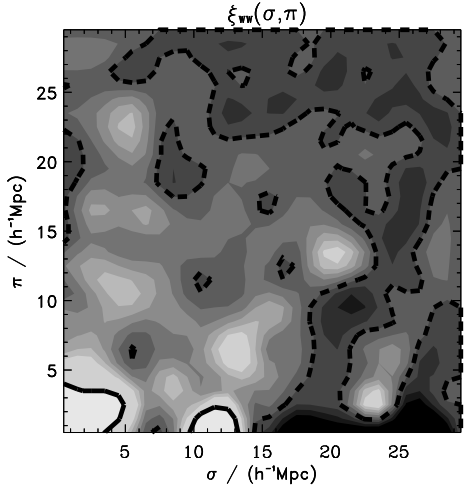

**Figure 13.** The redshift-space correlation function, $\xi(\sigma,\pi)$, of warm temperature-selected sample of QDOT galaxies. The contour levels are arithmetically spaced with a separation of 0.1: the dashed line is the contour for $\xi = 0.0$, while the solid line marks that for $\xi = 1.0$.

**Figure 14.** The redshift-space correlation function, $\xi(\sigma,\pi)$, of cool temperature-selected sample of QDOT galaxies. The contour levels are arithmetically spaced with a separation of 0.1: the dashed line is the contour for $\xi = 0.0$, while the solid line marks that for $\xi = 1.0$.

$$\chi^2 = \frac{1}{n} \sum_{i=1}^{n} \frac{(\rho_i - \rho_i')^2}{\lambda_i + \lambda_i'}, \quad (9)$$

where the density field, $\rho$ is calculated on a grid by $\rho_i = \sum_j w_j$, where the sum is over all galaxies in the subsample in the $i^{\rm th}$ cell, and each galaxy is given the unbiased weighting $w_j = 1/\phi(x_j)$. The mean density is normalised to unity. Each cell contributing to the reduced $\chi^2$ is weighted by the shot noise variance of the cell. The summation in equation (9) is taken over the $n$ non-empty cells containing both warm and cool IRAS galaxies.

To speed up calculation, we estimated the shot noise variance for each cell at a distance $r_i$ from the origin by integrating the appropriate weight over a Gaussian cell;

$$\lambda_i = 1 + (\sqrt{2\pi}R)^{-3} \int d^3 r\, w(r) \exp(-(\mathbf{r}-\mathbf{r}_i)^2/2R^2). \quad (10)$$

Here $R$ is chosen to match the size of a cubic cell. Assuming that the weighting scheme can be approximated as a power law across the cell, we find that

$$\lambda_i = 1 + \frac{w(r_i) u^{\alpha+1} e^{-u^2/2}}{\sqrt{2\pi}} \int_0^\infty dx\, x^{\alpha+1} e^{-x^2/2} (e^{xu} - e^{-xu}), (11)$$

where $u = r_i/R$, and $\alpha \equiv d\ln w(r)/d\ln r$. This last integral is related to the Parabolic Cylinder function, $D_\alpha(u)$, but in practise it is easier to calculate numerically. In principle, this variance should be equal to $1/\langle n\rangle_{\rm cell}$, where $\langle n\rangle_{\rm cell}$ is the expected galaxy occupancy for the $i^{\rm th}$ cell. However, this can be poorly determined due to the finite sampling of the cell, so instead we use the expected value for a homogeneous background.

If we assume that the $\chi^2$ difference has only Poissonian uncertainty, the variance on each estimate of $\chi^2$ can easily be evaluated. For samples with the same underlying mean density, we calculate that

$$\sigma_{\chi^2}^2 = \frac{1}{n^2} \sum_{i=1}^{n} \left[ \frac{(\lambda_i - 1)^3 + (\lambda_i' - 1)^3}{(\lambda_i + \lambda_i' - 2)^2} + 3 \right] - \frac{1}{n}. \qquad (12)$$

To proceed, we split the IRAS galaxies into warm and cool subsamples in such a way that both samples had the same selection function (see Section 3). This has the advantage of maximum possible overlap between the catalogues, and avoids the spurious effects of an underlying differential density gradient.

The density field for each sample was calculated over three sizes of cube, $R_{\rm box}$ = 4, 10 and 50 $\times$ 10$^3$kms$^{-1}$. Each cube was divided up into cells of size $r_{\rm cell} = R_{\rm box}/5$, $R_{\rm box}/10$, $R_{\rm box}/15$, and $R_{\rm box}/30$. This resulted in some overlap between the sizes of cells used to calculate the density, allowing us to check for consistency between differently sampled volumes. The choice of three sizes of cube allowed us to measure the density fields over a wide range of scales, without wasting time over sparsely sampled cells.

Fig. 12 shows the results of applying this test to the matched $n(z)$ QDOT samples. On scales larger than $10h^{-1}$Mpc, we see that the two density fields sample the same pattern in a manner consistent with the Poisson sampling hypothesis. Only five cells contribute to the value of $\chi^2$ for the smallest cell size ($r_{\rm cell} = 1.3h^{-1}$ Mpc), so the Poisson error bar assigned to that data point will be a serious underestimate of the true uncertainty, and so the highly significant difference between the redshift-space distributions of the warm and cool samples it purports to show should be regarded as spurious. The assumed Gaussian statistics should hold for the larger cell sizes, for which there are many more usable cells, so the significant deviation of the reduced $\chi^2$ statistic from unity for $2 \le r_{\rm cell}/h^{-1}$ Mpc $\le 10$, should be taken as direct evidence for a difference between the redshift-space clustering patterns of warm and cool IRAS galaxies: the origin of this difference is a tendency for smaller cells ($r_{\rm cell} \le 10h^{-1}$Mpc) to have larger numbers of warm galaxies than cool ones.

## 7 STRUCTURE IN THE REDSHIFT-SPACE CORRELATION FUNCTION

If, as our model and the results of the previous Section suggest, warm IRAS galaxies are preferentially located in richer environments than cool ones, we might expect to see a resultant difference in the structure in the redshift-space correlation functions, $\xi(\sigma, \pi)$, of the two samples: we look for such a difference in this Section.

We compute $\xi(\sigma, \pi)$ using the estimator

$$1 + \xi(\sigma, \pi) = \frac{D_A D_A(\sigma, \pi) \cdot R_A R_A(\sigma, \pi)}{[D_A R_A(\sigma, \pi)]^2}, \qquad (13)$$

with, as before, 20,000-particle random catalogues and a minimum variance weighting scheme, employing an appropriate model $J_3(s)$. We use separation bins of width 1.0 $h^{-1}$ Mpc in both directions, over the range 0 to $30h^{-1}$ Mpc. The raw $\xi(\sigma, \pi)$ data are very noisy, so we have smoothed them twice with a Shectman (1985) filter (a 1-2-1 boxcar in two dimensions) to produce the plots shown here.

In Figs. 13 and 14 we plot the smoothed $\xi(\sigma, \pi)$ results for the warm and cool temperature-selected samples. It is interesting to note that the structure in $\xi(\sigma, \pi)$ is far more pronounced in the plot for the warm sample than that for the cool sample: the same structures can often be seen in both plots, but with a higher amplitude in the warm plot, while in other cases, a strongly correlated region in the plot for the warm sample corresponds to an anti-correlated region in that for the cool sample. We show the corresponding plots for the warmer and cooler matched $n(z)$ samples in Figs. 15 and 16, where, again, the plot for the warmer sample exhibits more structure, albeit in a much less pronounced way than in Figs. 13 and 14. Figs. 13 – 16 together give some qualitative support to our model in which more warm than cool IRAS galaxies are found in richer environments, although, clearly, the increased concentration into richer environments is more a function of redshift than of temperature.

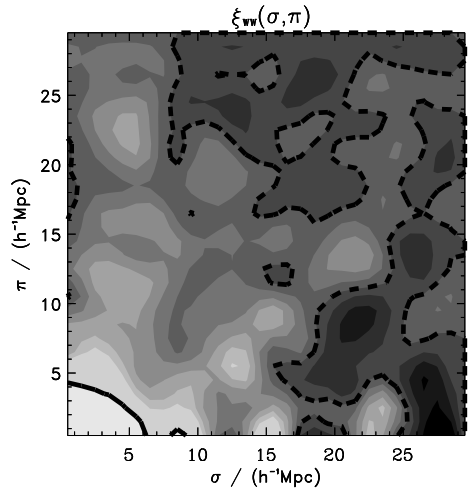

**Figure 15.** The redshift-space correlation function, $\xi(\sigma, \pi)$, of warmer matched $n(z)$ sample of QDOT galaxies. The contour levels are arithmetically spaced with a separation of 0.1: the dashed line is the contour for $\xi = 0.0$, while the solid line marks that for $\xi = 1.0$.

## 8 CROSS-CORRELATIONS WITH ABELL CLUSTERS

A direct way to probe whether warm IRAS galaxies are preferentially located in richer environments than cool ones would be to study the cross-correlations between the warm and cool samples and a suitable cluster catalogue. Unfortunately, if we were to restrict our warm and cool QDOT samples to the areas of the objective cluster catalogues drawn from the APM (Dalton et al. 1992) and EDSGC (Lumsden et al. 1992) galaxy catalogues we would have too few objects left to conduct the desired analysis. Since many Abell clusters (particularly those in the $R$=0 class) are suspected of being chance alignments of galaxies along the line of sight, rather than true rich clusters (e.g. Lumsden et al. 1992), our aim in calculating cross-correlations with Abell clusters in this Section is more modest: we simply wish to see whether the increasing strength of cross-correlations between Abell clusters and QDOT galaxies that Mo, Peacock & Xia (1993) see as they exclude QDOT galaxies with redshifts first $z \leq 0.02$ and then $z \leq 0.03$ can be accounted for by our temperature-dependent clustering model or whether it is a sampling effect.

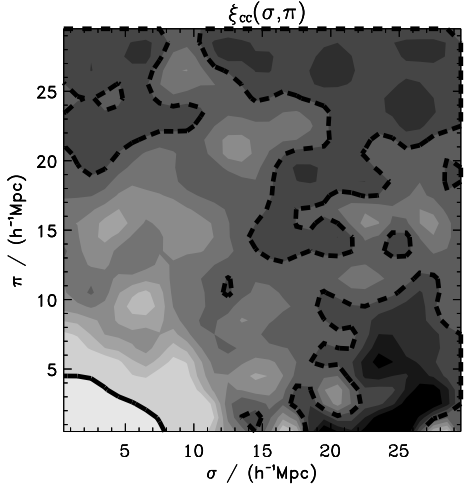

**Figure 16.** The redshift-space correlation function, $\xi(\sigma,\pi)$, of cooler matched $n(z)$ sample of QDOT galaxies. The contour levels are arithmetically spaced with a separation of 0.1: the dashed line is the contour for $\xi = 0.0$, while the solid line marks that for $\xi = 1.0$.

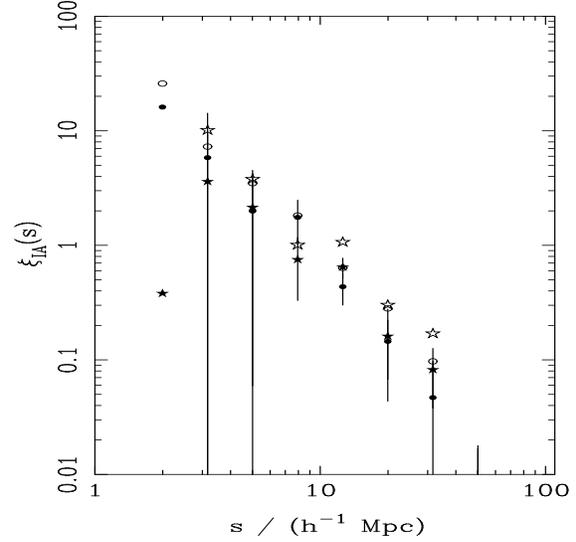

**Figure 18.** The cross-correlations of warm (stars) and cool (circles) temperature-selected samples of QDOT galaxies with Abell clusters: (a) the full $R \geq 0$ sample (filled symbols) and (b) the $R \geq 1$ subsample (open symbols).

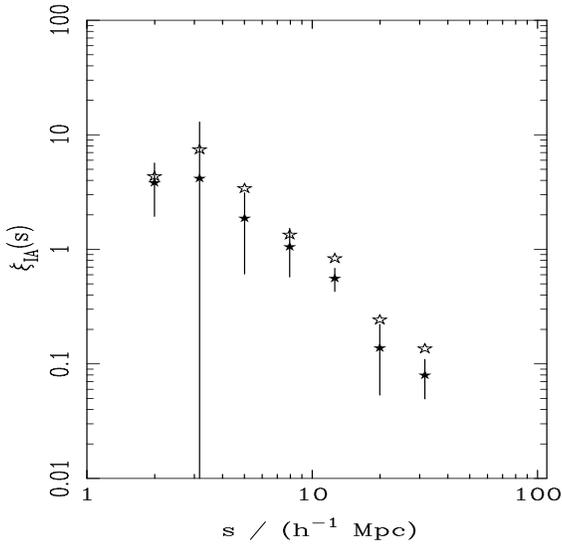

**Figure 17.** The cross-correlations of QDOT galaxies with Abell clusters: (a) the full $R \geq 0$ sample (filled symbols) and (b) the $R \geq 1$ subsample (open symbols).

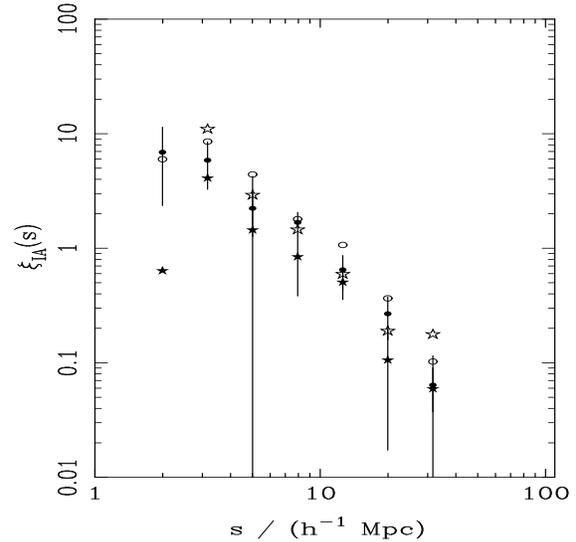

**Figure 19.** The cross-correlations of warmer (stars) and cooler (circles) matched $n(z)$ samples of QDOT galaxies with Abell clusters: (a) the full $R \geq 0$ sample (filled symbols) and (b) the $R \geq 1$ subsample (open symbols).

Our Abell cluster samples are drawn from those of Peacock & West (1992), with some updated redshift information (M. West, private communication). The Peacock & West (1992) sample covers the region of the sky at Galactic latitudes $|b| > 25°$, with the additional exclusion of the regions RA=$3^h - 6^h$, Dec.=$0° - 35°$ and RA=$15^h - 18^h$, Dec.=$-30° - 0°$, and covers the redshift range $0.01 < z < 0.08$. If we restrict this sample to the area not excluded by our IRAS mask, we obtain a volume-limited

Abell cluster sample covering 6.5 steradian, containing 220 $R = 0$ and 205 $R \geq 1$ clusters, which is about 90% complete in spectroscopic redshifts. The IRAS samples are reduced, too, by their restriction to the region allowed by Peacock & West (1992): the warm and cool temperature-selected samples fall to 691 and 409 galaxies, respectively, while the warmer and cooler matched samples now contain, respectively, 550 and 549 galaxies.

We compute the cross-correlation functions using the estimator

$$1 + \xi_{AB}(s) = \frac{D_A D_B(s)}{D_A R_B(s)} \cdot \frac{R_A R_B(s)}{R_A D_B(s)}, \quad (14)$$

where the meaning of the terms is clear from the discussion following equation (5). This is clearly a generalisation of the autocorrelation function estimator (equation 5) devised by Hamilton (1993) and, as we show in the Appendix, it shares with it the virtue of having an uncertainty which is second order in the uncertainties of the mean number densities of species A and B, as well as possessing the desirable quality of symmetry under the exchange of labels A and B. For these reasons we favour its use over that of the standard estimator

$$1 + \xi_{AB}(s) = \frac{D_A D_B(s)}{D_A R_B(s)} \cdot \frac{n_{R_A}}{n_A}, \quad (15)$$

which has an uncertainty that is first order in the uncertainty in the mean number density of B galaxies, as well as being asymmetric under exchange of A and B, which is aesthetically unpleasing in an estimator of an essentially symmetric quantity such as a cross-correlation function. As before, we use 20,000-particle random catalogues, employ the minimum variance weighting scheme with an appropriate $J_3(s)$ and compute error bars from the scatter between cross-correlation functions measured in octants.

In Fig. 17 we present the cross-correlation function for our full QDOT samples and our Abell cluster samples, both the full $R \geq 0$ sample and its $R \geq 1$ subsample. The best-fit power law for the cross-correlations with the $R \geq 1$ case over the range $3 \leq s \leq 30\ h^{-1}$ Mpc is $s_0 = 10.01 \pm 0.80\ h^{-1}$ Mpc, $\gamma = 2.04 \pm 0.23$, which is in good agreement with the results of Moore et al. (1994), who found $s_0 = 10.10 \pm 0.45\ h^{-1}$ Mpc, $\gamma = 1.75 \pm 0.10$ for slightly different galaxy and cluster samples. Figs. 18 and 19 present the corresponding plots for the warm and cool temperature-selected and warmer and cooler matched QDOT samples respectively. They clearly show that the clustering strength enhancement observed by Mo et al. (1993) must have its origin in a sampling effect, as the two pairs of warm and cool cross-correlation functions differ by less than $2\sigma$ over the full range ($1 \leq s \leq 100\ h^{-1}$ Mpc) studied.

## 9   DISCUSSION

In this paper we have questioned whether there might be a connection between the radial temperature and clustering strength gradients observed in the QDOT IRAS galaxy redshift survey. We showed how simple ideas of interaction-induced star formation in IRAS galaxies might produce such a connection and we proceeded to look for evidence for it, through a series of statistical tests of the clustering of warm and cool subsamples drawn from the QDOT redshift survey.

We obtained somewhat equivocal results. A significant difference was detected between the autocorrelation functions of the samples for separations $17 \leq s \leq 50\ h^{-1}$ Mpc, but this disappeared when we considered samples with matched redshift distributions, which suggests that the difference is caused by a sampling, rather than a temperature-dependent effect, given the similarity between the redshift distributions of the corresponding members of the temperature-selected and matched $n(z)$ pairs of samples. A counts-in-cells analysis found no difference between the cell count variances of warm and cool samples. In an attempt to improve upon the signal-to-noise levels obtained in these redshift-space clustering tests, particularly the last mentioned, we computed the projected cross-correlations of the warm and cool samples with the QIGC, from which the QDOT redshift survey was drawn. We found the warmer sample to have stronger cross-correlations on large scales, and this persisted, albeit only at a marginal significance, when we removed sampling effects by studying the matched $n(z)$ samples. We directly tested the null hypothesis that the galaxies in the warmer and cooler matched $n(z)$ samples are drawn from the same population in redshift-space and were able to rule it out on small scales, finding that cells of size $\leq 10 h^{-1}$ Mpc tended to have a significantly larger number of warmer galaxies than cooler ones. The structure in the $\xi(\sigma, \pi)$ plots for the warm and cool temperature-selected samples appeared to support the interpretation of these results as indicating that warmer IRAS galaxies are preferentially located in richer environments, in accordance with our interaction-induced model for the origin of their $60\mu m$ emission, but the corresponding plots for the matched $n(z)$ samples revealed that the greater concentration of warm galaxies in richer environments results more from the region of space sampled by the warm sample than from any directly temperature-dependent effect. We advocated the use of a new estimator for cross-correlation functions and applied it to the computation of the cross-correlations between our galaxy samples and a volume-limited sample of Abell clusters and this, too, suggested that the origin of the clustering strength gradient detected in QDOT by several previous authors lies in a sampling effect, rather than a temperature-dependence to the clustering of IRAS galaxies.

That significant sampling effects exist on the scale of the QDOT survey is, perhaps, surprising, especially since IRAS galaxies avoid rich clusters, so their clustering will be less sensitive to local variations in the cluster number density than that of optical galaxies: these effects should be taken into account when comparing theoretical predictions with clustering data from QDOT, especially on large scales. Only the higher signal-to-noise analysis possible with the PSC-$z$ IRAS redshift survey will reveal whether the temperature-dependent clustering component for which we claim a marginal detection here is of sufficient magnitude to warrant similar treatment. It serves, however, as an example of a type of effect that will becoming increasingly important in the study of large-scale structure in coming years. As the size of available redshift surveys increases by an order of magnitude it will become possible, and necessary, to consider in detail the clustering of subsamples of galaxies selected according to astrophysical criteria such as luminosity, morphology, spectral type, etc: the investigation of the coupling between astrophysical and clustering properties will not only be desirable for the correct comparison of observation and theory, but it may also shed invaluable light on the nature of bias, through the relationships between different types of galaxy and their environment.

The clustering analysis presented here can only indirectly test our interaction-induced star formation model for coupling the temperature and clustering strength gradients and, as we have seen, has produced only a marginal level of support for it. A more direct test can be made by quantitatively studying the environments of a large sample of IRAS galaxies selected over narrow ranges of redshift and $60\mu m$ luminosity, but with a wide range of $60\mu m$ to $100\mu m$ flux ratios: two of the present authors (RGM and ANT) will soon be reporting on exactly such a study (Goldschmidt et al., in preparation), which should reveal whether the link between interactions and enhanced far-infrared luminosity that exists for the most luminous IRAS galaxies continues down the $60\mu m$ luminosity function as far as the moderate luminosity galaxies that dominate redshift surveys of IRAS galaxies.

## 10 CONCLUSIONS

We have advanced a scenario in which the increase in clustering strength with redshift observed in flux-limited redshift surveys of IRAS galaxies arises as a result of the temperature gradients in the surveys, through a model in which the $60\mu m$ emission from IRAS galaxies has its origin in interaction-induced formation of massive stars. We have obtained some support for this scenario from clustering analyses of temperature-selected samples of QDOT galaxies, but we have shown that (at separations $\geq 10h^{-1}$ Mpc at least) the dominant component to the increase of clustering strength with redshift is a sampling effect, reflecting the local cosmography in the outer reaches the QDOT survey, which includes many rich clusters. This "cosmic variance" effect should be borne in mind by those using flux-limited samples of *IRAS* galaxies in large-scale structure studies.

## ACKNOWLEDGMENTS

We acknowledge financial support from SERC/PPARC, in the form of a research studentship and, latterly, a post-doctoral research assistantship on the QMW theory rolling grant, to RGM, an advanced fellowship to WS and a post-doctoral research assistantship to ANT. Phil Puxley, John Peacock and Alan Heavens are thanked for stimulating discussions, the QDOT team for access to their catalogue, Seb Oliver for the counts-in-cells software used in Section 5, and Mike West & John Peacock for communicating their Abell cluster samples and updated redshift information. We thank an anonymous referee for comments which helped to clarify the presentation of this paper.

## APPENDIX A: A NEW CROSS-CORRELATION FUNCTION ESTIMATOR

The standard estimator for the cross-correlation function, $\xi_{AB}(s)$, of galaxies of species A and B takes the form

$$1 + \xi_{AB}(s) = \frac{D_A D_B(s)}{D_A R_B(s)} \cdot \frac{n_{R_B}}{n_B}, \qquad (A1)$$

where the meaning of the terms follows from the paragraph below equation (5), of which equation (A1) is clearly the analogue. This estimator, like its analogous autocorrelation function estimator, suffers from having an uncertainty which is first order in the uncertainties in the mean number densities of the samples of A and B galaxies. It is also asymmetric under exchange of the labels A and B, which is clearly unattractive in an estimator for a symmetric beast like a cross-correlation function. These two facts combined can, in practice, produce the situation where equation (A1) and the estimator derived from it by the interchange of A and B can yield significantly different cross-correlations, even for the case of roughly equally-sized samples of A and B objects, where there is no *a priori* reason to favour one estimator over the other.

This undesirable situation may be avoided through the use of the following estimator:

$$1 + \xi_{AB}(s) = \frac{D_A D_B(s)}{D_A R_B(s)} \cdot \frac{R_A R_B(s)}{R_A D_B(s)}. \qquad (A2)$$

If A=B this clearly reduces to Hamilton's estimator for the autocorrelation function and, as we shall now show, it shares with it the property of having an uncertainty which is second order in the uncertainty in the mean number densities of the galaxy samples, as well as being manifestly symmetric under the exchange of labels A and B.

In the notation of Hamilton (1993), to which we refer the reader for the background to the following derivation and the notation in which it is written, equation (A2) becomes

$$1 + \xi_{AB}(s) = \frac{\langle N_A N_B \rangle}{\langle N_A W_B \rangle} \cdot \frac{\langle W_A W_B \rangle}{\langle W_A N_B \rangle}, \qquad (A3)$$

where $N_A$ denotes $N_{obs}^A$ galaxies of species A weighted in some fashion and $W_A$ is the "catalogue window", which is the selection function $\phi^A$ of A galaxies weighted in the same way as the A galaxies themselves: the angular brackets denote averages taken over all points in the catalogues with separations that place them in the bin centred on $s$.

Following Hamilton (1993), we define a pair window, $W_{12}^{AB}$, and a point window, $W_1^A$, as follows:

$$W_{12}^{AB} \equiv w_{12}^{AB} \phi_1^A \phi_2^B \qquad (A4)$$

and

$$W_1^A \equiv w_1^A \phi_1^A, \qquad (A5)$$

where $w_{12}^{AB}$ and $w_1^A$ are, respectively, pair and point weights: note that the pair weighting need not be separable and so $w_{12}^{AB}$ need not equal the product of the point weights $w_1^A$ and $w_2^B$. If the $n_1^A$ is the true space density of A galaxies at some point labelled 1, then, by the definition of the selection function, $N_{obs}^A = n^A \phi^A$ (neglecting a constant factor that cancels out in all the following analysis) and the quantities $N_A N_B$, $N_A W_B$ and $W_A W_B$ become

$$N_A N_B = w_{12}^{AB} N_{obs,1}^A N_{obs,2}^B = W_{12}^{AB} n_1^A n_2^B, \qquad (A6)$$

$$N_A N_B = w_{12}^{AB} N_{obs,1}^A \phi_2 = W_{12}^{AB} n_1^A, \qquad (A7)$$

$$W_A W_B = w_{12}^{AB} \phi_1^A \phi_2^B = W_{12}^{AB}. \qquad (A8)$$

It follows, therefore, that equation (A3) may be written as

$$1 + \xi_{AB}(s) = \frac{\langle W_{12}^{AB} n_1^A n_2^B \rangle \langle W_{12}^{AB} \rangle}{\langle W_{12}^{AB} n_1^A \rangle \langle W_{12}^{AB} n_2^B \rangle}. \qquad (A9)$$

Now, if we define $\delta_1^A$ to be the true overdensity in A galaxies at point 1 and $\bar{n}^A$ to be their true mean density, then equation (A9) becomes

$$1 + \xi_{AB}(s) = \frac{\langle W_{12}^{AB} \bar{n}^A \bar{n}^B (1+\delta_1^A)(1+\delta_2^B) \rangle \langle W_{12}^{AB} \rangle}{\langle W_{12}^{AB} \bar{n}^A (1+\delta_1^A) \rangle \langle W_{12}^{AB} \bar{n}^B (1+\delta_2^B) \rangle}. \qquad (A10)$$

If we then follow Hamilton (1993) in defining the cross-correlation function between the A galaxies and the B catalogue to be

$$\psi_{12}^{AB} \equiv \frac{\langle W_{12}^{AB} \delta_1^A \rangle}{\langle W_{12}^{AB} \rangle}, \qquad (A11)$$

and the windowed cross-correlation function of A and B galaxies (which is the best estimate of the true cross-correlation function obtainable from the galaxy samples), to be

$$\hat{\xi}_{12}^{AB} \equiv \frac{\langle W_{12}^{AB} \delta_1^A \delta_2^B \rangle}{\langle W_{12}^{AB} \rangle}, \qquad (A12)$$

we see that the estimated cross-correlation function reduces to the form

$$\xi_{AB}(s) = \frac{\hat{\xi}^{AB}(s) - \psi^A(s)\psi^B(s)}{[1+\psi^A(s)][1+\psi^B(s)]}. \qquad (A13)$$

Now, since (Hamilton 1993) $\psi^A \sim \mathcal{O}(\bar{\delta}^A)$, where $\bar{\delta}^A \equiv \langle W_1^A \delta_1^A \rangle / \langle W_1^A \rangle$ is the mean overdensity of A galaxies in the catalogue, it follows that the uncertainty in $\xi_{AB}(s)$ is second order in the uncertainties in the mean densities, which is better than the standard estimator, which would become, in this notation,

$$\xi_{AB}(s) = \frac{\hat{\xi}_{AB}(s) + [\psi^B(s) - \bar{\delta}^B] - \psi^A(s)\bar{\delta}^B}{[1+\psi^A(s)](1+\bar{\delta}^B)}, \qquad (A14)$$

which is first order in $\bar{\delta}^B$.